\documentclass[pre,twocolumn,floatfix,aps,showpacs]{revtex4}
\usepackage{graphicx}
\usepackage{amsmath}
\usepackage{amsfonts}
\usepackage{amssymb}
\usepackage{epsf}
\usepackage{epsfig}
\usepackage{bm}

\begin{document}
\title{An analytical approach to sorting in periodic potentials}
\author{James P. Gleeson$^1$, J. M. Sancho$^2$, A. M. Lacasta$^3$, and
Katja Lindenberg$^4$} \affiliation{ $^{(1)}$ Department of Applied
Mathematics, University College Cork, Cork,
Ireland\\
$^{(2)}$
Departament d'Estructura i Constituents de la Mat\`eria,
Facultat de F\'{\i}sica, Universitat de Barcelona,
Diagonal 647, E-08028 Barcelona, Spain\\
$^{(3)}$
Departament de F\'{\i}sica Aplicada,
Universitat Polit\`{e}cnica de Catalunya,
Avinguda Doctor Mara\~{n}on 44, E-08028 Barcelona, Spain\\
$^{(4)}$
Department of Chemistry and Biochemistry 0340 and Institute for
Nonlinear Science,
University of California, San Diego, La Jolla, California 92093-0340,
USA\\
}
\date{\today}
\def\pv{{\mathbf{p}}}
\def\qv{{\mathbf{q}}}
\def\rv{{\mathbf{r}}}
\def\xv{{\mathbf{x}}}
\def\kv{{\mathbf{k}}}
\def\yv{{\mathbf{y}}}
% add p3.tex definitions:
\def\xv{{\mathbf{x}}}
\def\rv{{\mathbf{r}}}
\def\ev{\boldsymbol{ \eta}}
\def\t{\phi}
\def\xvz{{\mathbf{x_0}}}
\def\rvz{{\mathbf{r_0}}}
\def\zvz{{\mathbf{z_0}}}
\def\rv{{\mathbf{r}}}
\def\uv{{\mathbf{u}}}
\def\yv{{\mathbf{y}}}
\def\lb{\left(}  %left bracket
\def\rb{\right)}
\def\d{\delta} % delta
\def\p{\partial}   %partial
\def\Lc{\mathcal{L}}  % L curly
\def\gv{\mathbf{g}}  % g vector
\def\g{\gamma}
\def\a{\alpha}
\def\b{\beta}
\def\Sc{\mathcal{S}}   % curly S
\def\kv{{\mathbf{k}}}
\def\jv{{\mathbf{j}}}
\def\xv{{\mathbf{x}}}
\def\Cc{\mathcal{C}}
\def\be{\begin{equation}}
\def\h{\frac{1}{2}}  % half
\def\q{}
\def\f{\frac}
\def\w{\omega}
\def\doubleu{\omega}
\def\doubleuv{\boldsymbol{ \omega}}
\def\Qv{\mathbf{Q}}
\def\pr{\rho}
\def\eps{\epsilon}
\def\th{\theta}
\def\zv{{\mathbf{0}}}
\def\Fv{{\mathbf{F}}}
\def\vv{{\mathbf{v}}}
\def\T{\mathcal{T}}
\begin{abstract}
There has been a recent revolution in the ability to manipulate 
micrometer-sized objects on surfaces patterned by traps or obstacles 
of controllable configurations and shapes.  One application of this
technology is to separate particles driven across such a surface by an
external force according to some particle characteristic such as size or
index of refraction.  The surface features cause the trajectories of
particles driven across the surface to deviate from the direction of the
force by an amount that depends on the particular characteristic, thus
leading to sorting.  While models of this behavior have provided a good
understanding of these observations, the solutions have so far been
primarily numerical.  In this paper we provide analytic predictions for
the dependence of the angle between the direction of motion and the
external force on a number of model parameters for periodic as well as
random surfaces.  We test these predictions against exact numerical
simulations.  
\end{abstract}


\maketitle

\section{Introduction}
\label{introduction} Transport of particles driven by external
forces across modulated potential surfaces, and the associated
ability to sort mixtures of particles according to some feature
sensitive to this modulation, has attracted considerable recent
interest~\cite{Duke,Reichhardt,Korda,Grier,Gopinathan,McDonald,Huang,Ladvac,Pelton,PRL,NewJPhys,SPIE}.
The surface modulation may be periodic or random, and it may
consist of traps, obstacles, or a combination of both.  Typical
sorting parameters include particle size, particle index of
refraction, and particle mass.  The underlying mechanism in all
cases relies on the fact that while the particles attempt to
follow the external force, the traps or obstacles induce
systematic deviations in the trajectories so that there is a
non-zero average angle between the trajectory and the external
force.  If this non-zero angle depends on some characteristic of
the particle such as size, then particles that differ in this
characteristic emerge at different angles and can therefore be
sorted in a very effective way.

There are a number of experimental demonstrations of these
capabilities~\cite{Duke,Korda,Grier,McDonald,Huang,Ladvac}, and
there are also fairly extensive numerical simulations of various
model formulations~\cite{Reichhardt,Gopinathan,PRL,NewJPhys,SPIE}
that are quite successful in explaining the experimental
phenomena.  These simulations to some extent clarify the roles of
a number of physical variables such as, for example, the direction
of the external force, surface geometry, and ambient temperature,
that affect the experimental outcomes.  On the other hand,
analytic results are relatively rare~\cite{Pelton}, and yet it is
undoubtedly useful to have predictive analytic formulas,
especially with a view to optimizing the sorting mechanism. In
this paper we take a step in this direction, deriving approximate
results that are shown to reproduce some of the important earlier
numerical results.  While our results are limited to certain
parameter regimes, we believe that they may prove useful in
designing experimental sorting potentials and in identifying
likely regions of parameter space for more detailed numerical
study.

In Sec.~\ref{model} we describe the model and define the
quantities to be calculated, specifically, the angle between the
average velocity of the particles and the external force. In our
discussions we include periodic potentials as well as random
potentials. In Sec.~\ref{approximate} (accompanied by an Appendix)
we present approximate equations for the average velocity valid for
sufficiently strong forces and/or high temperatures, and discuss
how these approximations might be continued to further orders than
those retained here. In Sec.~\ref{resultsp} we compare our
theoretical results with those of numerical simulations for
periodic potentials, and in the course of this comparison we
refine our approximations so as to eliminate an unphysical
outcome. With this adjustment we improve the theoretical
predictions. In Sec.~\ref{resultsr} we carry out a similar
comparison for random potentials, and again find very good
agreement between theory and numerical simulations.  We summarize
our results and indicate some future directions in
Sec.~\ref{conclusion}.

\section{The model}
\label{model}
We consider identical non-interacting particles moving across a
surface described
by a two-dimensional potential $V(x,y)$ of unit height or depth and unit
period in both directions.
The equation of motion for
the particles is given by a Langevin equation for each component of the
particle displacement~\cite{PRL},
\begin{eqnarray}
\dot x &=& -\frac{\p}{\p x}V(x,y)+F \cos \th+\sqrt{2 \mathcal{T}}
\xi_x(t) \nonumber\\
\dot y &=& -\frac{\p}{\p y}V(x,y)+F \sin \th+\sqrt{2 \mathcal{T}}
\xi_y(t) \label{1}.
\end{eqnarray}
The dots denote time derivatives, $\mathcal{T}$ is the
dimensionless temperature, and the thermal fluctuation terms
$\xi_i(t)$ are Gaussian and $\delta$-correlated,
\begin{equation}
\left< \xi_i(t)\xi_j\left(t^\prime\right)\right>= \delta_{i
j}\delta\left(t-t^\prime\right).
\end{equation}
Note that the equations are in dimensionless form, with the scaled
temperature $\T$ and force magnitude $F$ given in terms of
physical parameters as in Eq.~(4) of \cite{PRL}. The system is
assumed to be overdamped so that inertial terms $\ddot{x}$ and
$\ddot{y}$ have been dropped from the equations of motion. The
constant external force vector is
\begin{equation}
\mathbf{F} = F \cos \th \,\mathbf{i}+F \sin\th\, \mathbf{j},
\end{equation}
and all particles are assumed to start at the origin.

Our goal is to calculate the direction and magnitude
of the velocity of the particles averaged over the thermal
fluctuations. The long-time limit of this velocity is denoted by
$\left<\vv\right>$, and is decomposed into Cartesian components as
\begin{equation}
\left<\vv\right> = \left<v_x\right> \mathbf{i}+ \left<v_y\right>
\mathbf{j}.
\end{equation}
The angle brackets denote averaging over the thermal noise.
If the potential $V(x,y)$ is random, then we also perform an
average over realizations of the potential, denoting such an
ensemble average by an overbar: $\overline{\left<\vv\right>}$. To
calculate the average velocity for a periodic potential, we write
the Fokker-Planck equation for the concentration $c(\xv,t)$ of the
particles obeying the equations of motion (\ref{1}),
\begin{equation}
\frac{\p c}{\p t}+\nabla\cdot\left[\left(\Fv-\nabla
V\right)c\right] - \T \nabla^2 c =0, \label{FPE}
\end{equation}
with initial condition $c(\xv,0)=\delta(\xv)$. If this equation
could be solved for the concentration $c(\xv,t)$ at time $t$, then
the desired average velocity vector follows from
\begin{equation}
\left<\vv\right> = \lim_{t\rightarrow \infty}\left<\frac{d \xv}{d
t}\right> = \lim_{t\rightarrow \infty} \frac{d}{d t}\int d\xv\,
\xv \, c(\xv,t). \label{vel}
\end{equation}
While an exact solution of Eq.~(\ref{FPE}) is not in general
possible, an approximate solution for the concentration may be
derived using a number of methods.  We implement such an
approximate solution in the next section, and also generalize the
results to the case of a random potential by averaging over the
disorder to obtain $\overline{\left<\vv\right>}$.

Of particular interest in sorting applications is the deflection
angle $\a$ of the velocity from the external force direction,
given by the relation~\cite{PRL}
\begin{equation}
\tan \a = \frac{\left< v_\perp\right>}{\left< v_\parallel\right>}
= \frac{-\left<v_x\right> \sin\th+\left<v_y\right>
\cos\th}{\left<v_x\right> \cos\th+\left<v_y\right> \sin\th}.
\label{alphaeqn}
\end{equation}
(If the potential is random, each velocity component
$\left<v_\perp\right>$, $\left<v_x\right>$ etc. in this formula
should be replaced by its average over the disorder:
$\overline{\left<v_\perp\right>}$, $\overline{\left<v_x\right>}$
etc.). We will explore the behavior of this angle through that of
each component $\left<v_x\right>$ and $\left<v_y\right>$ of the
average velocity, and examine the accuracy of our
approximations in capturing these behaviors.

To avoid overly complicated formulas, we adopt the convention
throughout the remainder of this paper (except in the Appendix) that
the notation $\vv$ (or its components $v_x$, $v_y$, $v_\perp$,
$v_\parallel$) stands for the averaged quantity $\left<\vv\right>$
(or, respectively, $\left<{v_x}\right>$, $\left<{v_y}\right>$,
$\left<{v_\perp}\right>$, $\left<{v_\parallel}\right>$). If the
potential is random, $\vv$ (or its components) stands for
$\overline{\left<\vv\right>}$ (or its respective averaged
components). Averages of all other quantities will be denoted
explicitly using angle brackets and overbars as appropriate.

\section{Approximate solutions}
\label{approximate}

The experimentally interesting regimes involve an external force
of magnitude sufficiently larger than the well depth ($F > 1$) so
that particles do
not easily become trapped.  If a particle did become trapped, it would
have to be extracted by a sufficiently large thermal fluctuation to
continue moving across the surface.  The particles of interest
in a sorting context are those that do not become trapped during the
course of the experiment.
On the other hand, if the force is too large, it simply drags
the particles along, the features of the potential become essentially
invisible, and particle separation, which relies on the effect of the
potential on the particle trajectories, does not occur.
At the same time, the temperature of the system should not be so high
as to obliterate the features of the potential. A temperature that is
too high would again lead to an uninteresting situation in
that particle separation would again not be observed.  The regime of
interest is thus that of an external force that is large but not too
large, and a temperature that is as low as possible in theory and in
practice.

\subsection{First order approximation for periodic potentials}
\label{firstorder}
In the Appendix we show how a systematic large force and/or high
temperature approximation may be obtained.  The first-order
approximation leads to the average velocity vector
\begin{equation}
\vv = \Fv+\frac{i}{(2\pi)^4}\int d\kv \, \frac{k^2 \kv}{\T
k^2-i\kv\cdot\Fv} \, \hat Q(\kv). \label{formula}
\end{equation}
where for periodic potentials the function $\hat Q(\kv)$ is
defined by
\begin{equation}
\hat Q(\kv) = \hat V(\kv) \hat V(-\kv) .
\end{equation}
Here $k^2=\kv.\kv$ and $\hat V$ is the Fourier transform of the
potential $V(\xv)$ as defined in Eq.~(\ref{A2}).

For later comparisons, it is useful to exhibit the result of this
approximation for the simplest one-dimensional version of our
problem, one with a simple cosinusoidal potential. This case has
been studied extensively (see, for example, Chapter 11
of~\cite{Risken}), and exact solutions may be found for the
average velocity. The equation of motion is
\begin{equation}
\frac{d x}{d t} =U + 2 \pi \sin(2\pi x)+\sqrt{2 \T} \xi(t). \label{1r}
\end{equation}
We call the forcing $U$ instead of $F$ because for the simplest
separable potential in two dimensions
we will be able to identify the average speed
$v(U)=\left<\dot x\right>$
with the components of the two-dimensional velocity via the relations
$v_x= v(F\cos\th)$ and $v_y=v( F \sin\th)$, thus
providing a way to determine the deflection angle in a two-dimensional
case via a one-dimensional calculation.  The result is
\begin{equation}
v(U) \approx U -  \frac{2 \pi^2 U}{4\pi^2 \T^2 + U^2}.
\label{oned}
\end{equation}
A number of points are noteworthy.  Firstly, a simple sinusoidal potential
has minima (which we think of as traps) as well as maxima (which we think
of as barriers) relative to a flat landscape~\cite{NewJPhys}.
Secondly, we observe that
the net effect of the potential in one dimension is to slow down the particles.
Thirdly, when the associations
$v_x= v(F\cos\th)$ and $v_y=v( F \sin\th)$ are appropriate, we can
immediately see that the deflection $\alpha$ is associated with the fact
that the ``slowing down" is different for each component for most angles
$\th$.

The technique yielding the approximate formula (\ref{formula}) in
the case of a periodic potential may readily be extended to
random potentials, as simulated numerically
in~\cite{PRE04}, for instance. Indeed, the method of deriving
Eq.~(\ref{formula}) was originally developed for the calculation
of particle concentrations when advected by potentials which are
random in space~\cite{BouchaudGeorges} and also in
time~\cite{PhysFlu,PullinPhysFlu}. The possibility of sorting on
random potentials can thus be examined using analytic
approximations.

\subsection{First order approximation for random potentials}
The appropriate kernel in formula (\ref{formula}) for a random
potential is
\begin{equation}
\hat Q(\kv) = 4 \pi^2 \hat E(\kv) \label{randQ}
\end{equation}
where $\hat{E}(\kv)$ is the ``energy spectrum'' of the potential.
The energy spectrum is defined as the Fourier transform of the
(disorder-averaged) correlation function of the potential
$\overline{V\left(\xv^\prime\right)V\left(\xv^\prime+\xv\right)}$.
The disorder is assumed to be homogeneous, so that the correlation
function depends only on the difference vector $\xv$; recall that
the overbar denotes averaging over the ensemble of random
potentials.

It is again useful to exhibit the first order approximation for the
simplest one-dimensional version of the problem.  The equation of motion
is
\begin{equation}
\frac{d x}{d t} =U - \frac{\partial}{\partial x} V(x)
+\sqrt{2 \T} \xi(t),
\label{1Drand}
\end{equation}
where $V(x)$ is now a zero-mean random modulation with spectrum
$\hat E(k)$.
The one-dimensional velocity averaged over realizations of the
noise $\xi(t)$ and of the random potential is then approximately
given by
\begin{equation}
v(U)=U-\frac{U}{2\pi}\int_{-\infty}^\infty dk\, \frac{k^2
\hat{E}(k)}{\T^2 k^2+  U^2}. \label{1DrandFDC1}
\end{equation}
A more explicit result requires the specification of the correlation
function.

It is an interesting general result that regardless of the
specific form of the correlation function, the average deflection
angle is zero when the random potential in our two-dimensional
scenario is isotropic, that is, when the correlation function
$\overline{V\left(\xv^\prime\right)V\left(\xv^\prime+\xv\right)}$
depends only on the magnitude $|\xv|$ of the difference vector and
not on its direction (see, for example, Eq.~(14) of \cite{PRE04}).
In this case the energy spectrum $\hat{E}(\kv)$ depends only on
the magnitude $k=|\kv|$ of its argument. Using (\ref{randQ}), we
consider the wave-vector integral in (\ref{formula}) by writing
$\kv$ in terms of a component $k_\parallel=k\cos\phi$ parallel to
the force $\Fv$, and a component $k_\perp=k\sin\phi$ perpendicular
to $\Fv$, where $\phi$ is the angle between $\kv$ and $\Fv$. With
the two-dimensional integral written in polar coordinates,
the average perpendicular velocity is found to be
\begin{equation}
v_\perp =\frac{i}{(2\pi)^2}\int_0^{\infty} dk  \, k
\int_0^{2\pi}d\phi\, \frac{k^3 \sin\phi}{\T k^2-i k F \cos\phi} \,
\hat{E}(k). \label{randvperp}
\end{equation}
The isotropy of the potential means that the energy spectrum is
independent of the angle $\phi$, and therefore the integration
over $\phi$ may be performed in (\ref{randvperp}), giving the
result $v_\perp = 0$. We conclude that for isotropic random
potentials the deflection angle is zero. This means that sorting
by deflection angle is not possible in isotropic random
potentials.

\subsection{Higher order approximations}
\label{secondordersec} It is useful to explore the behavior and
magnitude of the second and higher order approximations for the
velocity.  We do not do this in full generality, but only for the
simplest one-dimensional problems~(\ref{1r}) and (\ref{1Drand}).

Following the procedure
outlined in the Appendix, a second iteration
is somewhat cumbersome but straightforward, and leads in the periodic
case to the approximate velocity
\begin{eqnarray}
v(U) &\approx U & - \frac{2 \pi^2 U}{4\pi^2 \T^2 + U^2}
\nonumber \\
&&\hspace{-0.5cm}- 2\pi^4 \frac{U^3-20 \pi^2 U\T^2}{(U^2+4\pi^2
\T^2)^2(U^2+16 \pi^2
  \T^2)}.
\label{higher}
\end{eqnarray}
The last term is the second order correction.
In a subsequent section, where we make comparisons with exact
results, we comment on some higher order terms in this series.
Here we simply point out that for $\T=0$ the full infinite series
sums to the known exact solution~\cite{Risken} $v(U)/U =
\sqrt{1-4\pi^2/U^2}$.

In the case of a one-dimensional random potential, continuing the
iteration to the next order extends Eq.~(\ref{1DrandFDC1}) to
\begin{eqnarray}
v(U)&=&U-\frac{U}{2\pi}\int_{-\infty}^\infty dk\, \frac{k^2
\hat{E}(k)}{\T^2 k^2+  U^2} \nonumber\\
&& + \frac{1}{(2\pi)^2}\int_{-\infty}^\infty \!
dp\!\int_{-\infty}^\infty \!dq\,  p^2 \hat{E}(p)\, q^2
\hat{E}(q) \nonumber\\
&& \times \left[\frac{i}{(i U-\T p)(i
U-\T(p+q))(i U-\T p)} \right.\nonumber\\
&&\left.+\frac{i}{(i U-\T p)(i U-\T(p+q))(i U-\T
q)}\right].
\nonumber\\
&&
\label{secondorder}
\end{eqnarray}

Finally, while we recognize that the approximation methods based
on the truncation of a series precludes a straightforward
determination of the regimes of validity, one might expect that
high temperature and/or strong external forcing ($\T \gg 1$ and/or
$F\gg 1$) would be sufficient to make the approximations presented
in this section reasonably accurate. As we shall see later, such
an assertion requires some caveats, but numerical simulation
results help resolve the issues which arise.

\section{Results for periodic potentials}
\label{resultsp} In this section we test the approximate results
against exact simulation results for two-dimensional periodic
potentials, explain why large $F$ is not necessarily sufficient
for agreement at the orders developed in the last section, and
modify our approximations nonperturbatively so as to correct for
the source of disagreement.

If the potential is 1-periodic and even in both $x$ and $y$, such as
the one used in~\cite{PRL}, it may be expanded in a Fourier
series as
\begin{equation}
V(\xv)=\sum_{n=0}^M \sum_{m=0}^M a_{n m} \cos(2\pi n x)\cos(2 \pi
m y), \label{FSpot}
\end{equation}
where $M$ may be infinite, although in our calculations we
retain only a finite number of modes in each
direction. After some algebra, Eq.~(\ref{formula}) yields
explicit expressions for the average velocity in the $x$ and $y$
directions as sums over Fourier modes,
\begin{eqnarray}
v_x &=& F \cos\th - \frac{\pi^2}{2} \sum_{n=0}^M \sum_{m=0}^M a_{n
m}^2 d_n d_m F n (n^2+m^2) \nonumber\\
&& \times \left[ \frac{n \cos \th-m \sin
\th}{4\pi^2 \T^2(n^2+m^2)^2+F^2(n \cos\th-m \sin\th)^2} \right.
\nonumber
\\
&&\left. +\frac{n \cos \th+m \sin \th}{4\pi^2
\T^2(n^2+m^2)^2+F^2(n
\cos\th+m \sin\th)^2}\right] \nonumber \\
v_y &=& F \sin\th - \frac{\pi^2}{2} \sum_{n=0}^M \sum_{m=0}^M a_{n
m}^2 d_n d_m F m (n^2+m^2) \nonumber\\
&& \times \left[ \frac{m \sin \th - n \cos
\th}{4\pi^2 \T^2(n^2+m^2)^2+F^2(n \cos\th-m \sin\th)^2}\right.
\nonumber
\\
&&\left. +\frac{m \sin \th+n \cos \th}{4\pi^2
\T^2(n^2+m^2)^2+F^2(n \cos\th+m \sin\th)^2}\right], \nonumber\\
&&
\label{fullformulas}
\end{eqnarray}
where the coefficients $a_{nm}$ depend on the specific potential, and
\begin{equation}
d_n=1+\delta_{n0}=\left\{ \begin{array}{cl}
                            2 & \text{ if }n=0\\
                            1 & \text{ if }n>0.
                            \end{array}
\right. \label{deqn}
\end{equation}
Note that to this order of approximation the Fourier coefficients
$a_{nm}$ appear only as $a^2$, so
that the resulting velocities are the same regardless of the overall
sign of the potential, e.g., whether the potential consists of traps or
of obstacles.

To examine the usefulness of the approximate formulas
(\ref{fullformulas}), we choose the periodic potential used in
\cite{PRL},
\begin{equation}
V(x,y) = \frac{-1}{1+e^{-g(x,y)}}, \label{actualpot}
\end{equation}
with $g(x,y)=5\left[ \cos(2\pi x)+\cos(2\pi y)-2 B\right]$, and
with various values of the parameter $B$ (associated with
different particle sizes~\cite{PRL}). The negative sign in
the numerator indicates
that we are considering traps (but, as noted above, this sign does not
matter for the assessment of the validity of the approximation).
For each value of $B$, the
potential is first expressed in terms of its Fourier series
(\ref{FSpot}) to determine the Fourier coefficients $a_{n m}$.
\begin{figure}
\centering
 \epsfig{figure=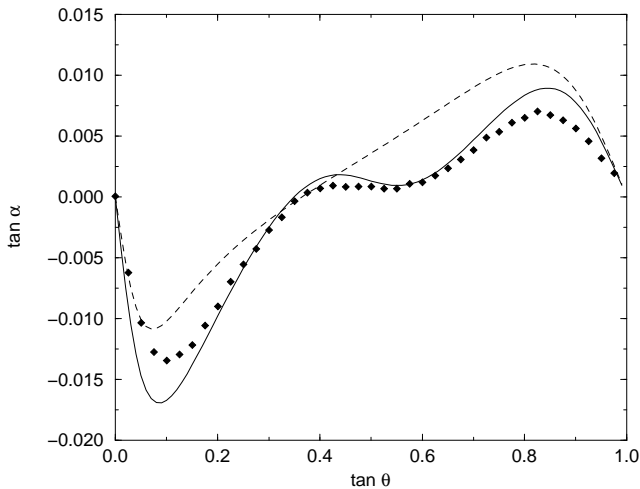,width=8.5cm}
\caption{ The dependence of the deflection angle $\a$ on the
forcing angle $\th$ at temperature $\T=0.1$, and for parameters
$F=8$ and $B=0.9$. Numerical simulation results (solid diamonds)
agree well with theoretical predictions using $M=6$ modes (solid
line) in the Fourier series (\ref{FSpot}). For comparison, the
theoretical prediction using only $M=1$ modes is also shown
(dashed line). } \label{fig1}
\end{figure}

Figure~\ref{fig1} shows the angle $\tan \a$ defined in
Eq.~(\ref{alphaeqn}) plotted against $\tan \th$ for $F=8$ and
$\T=0.1$, and with $B=0.9$. Note the degree of agreement between
the exact and approximate results, especially the positive and
negative regions of deflection angle which reflect the ``terrace
phenomenon'' or ``locked-in'' states observed in
experiments~\cite{Duke,Korda,Grier,McDonald,Huang,Ladvac}. Note
also the accord in the order of magnitude of the deflection. These
indications point to the qualitatively satisfactory performance of
the approximate formula (\ref{formula}) even at the low
temperature $\T=0.1$.
Also of interest here is the dependence of the results on the
number $M$ of Fourier modes retained in the expansion
(\ref{FSpot}) of the potential. The shape of the potential
(\ref{actualpot}) is well approximated with $M=6$; however our
results with $M$ as low as $1$ show that the correct order of
magnitude of the deflection can be predicted using only the lowest
Fourier harmonics of the potential.

The Fourier decomposition provides an opportunity to understand
the role of various symmetry contributions to the sorting
capability of the surface.  For this purpose, we examine two
special cases with $M=1$.  The potential
\begin{equation}
V_a(x,y)=\cos(2 \pi x)+\cos(2 \pi y) \label{separable}
\end{equation}
corresponds to having $a_{0 1}=a_{1 0}=1$, with all other $a_{n
m}$ being zero. This separable potential was used in
\cite{NewJPhys} and recently in \cite{JPhysCondMat}, and we
discuss it further subsequently. For the parameters used in
Fig.~\ref{fig1}, the potential $V_a$ yields deflection angles
which are negative for all forcing directions.
In contrast, the potential
\begin{equation}
V_b(x,y) = \cos(2\pi x)\cos(2 \pi y),
\end{equation}
corresponding to $a_{1 1}=1$, with $a_{n m}=0$ otherwise, gives
positive deflection angles  for all forcing directions. It is the
(appropriately weighted) combination of the potentials $V_a$ and
$V_b$ in the Fourier series of (\ref{actualpot}) which generates
the crossover from negative to positive deflection angle in the
$M=1$ curve of Fig.~\ref{fig1}.

In Fig.~\ref{chopping} we show results of numerical simulations of
sorting in the separable potential $V_a$ at temperature $\T=0.1$
and for various magnitudes of the forcing $F$. As noted following
Eq.~(\ref{1r}), separable potentials allow the application of
one-dimensional results to the two-dimensional case by setting
$v_x=v(F \cos\th)$, $v_y=v(F \sin\th)$. The dashed curves in
Fig.~\ref{chopping} show the deflection angle predicted using the
first order approximation (\ref{oned}).
While the agreement is quite good at the higher angles $\theta$, it is
clearly not quantitatively satisfactory at low angles.  The second
order approximation from Eq.~(\ref{higher})  yields the solid
curves, and further improves the agreement at the higher values of
$\theta$ and pushes this agreement toward lower $\theta$, but does
not greatly improve the low-$\theta$ situation. The difficulty can
easily be traced in the case of the separable potential and, by
inference, for more complex potentials, by recalling that $v_y=v(
F \sin\th)$. Clearly, at low angles the argument of $v_y$ is
necessarily small no matter the magnitude of $F$.
Equations~(\ref{oned}) and (\ref{secondorder}) make it clear that
at sufficiently small argument the one-dimensional approximate
velocity becomes negative. This unphysical result is reflected in
the misbehavior of $v_y$ at small $\th$ and leads to the
disagreements observed at small angles.

To test this hypothesis, we introduce an enhanced approximation,
which consists of calculating $v_x$ and $v_y$ from the first-order
or second-order iteration approximation, as we have done so far,
but then replacing $v_y$ by max$[v_y,0]$ and $v_x$ by
max$[v_x,0]$.  This eliminates the unphysical negative values of
the components, replacing them by zero, without affecting positive
outcomes. We shall call this the {\em adjusted truncation}, and
its effect on the predictions is to cut off the curves to the left
of the dotted line $\tan\alpha=-\tan\th$ in Fig.~\ref{chopping}.
The agreement between the adjusted truncation and the numerical
results is clearly very good , with the remaining differences
arising from the abrupt replacement of negative velocities by zero
velocities in the adjusted truncation (and, for the second order
case, a very small region near $\theta=0$, where a positive
deflection angle is erroneously predicted).

\begin{figure}
\begin{center}
\epsfig{figure=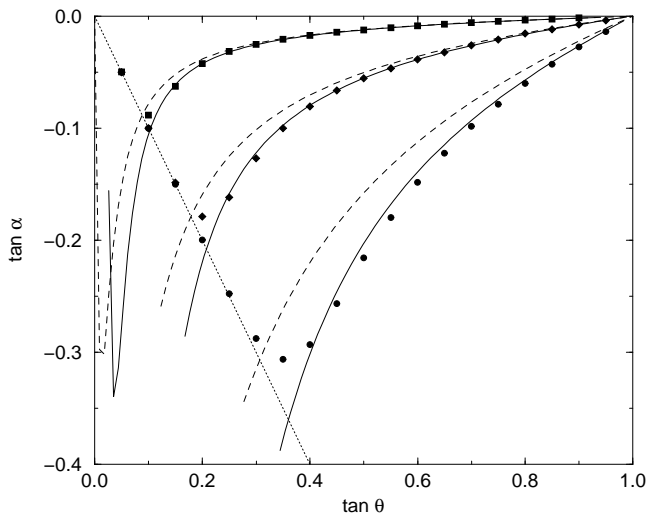, width=8.5cm}
\end{center}
\caption{Deflection angle for separable two-dimensional potential
(\ref{separable}) with $\T=0.1$ and $F=15$, $25$, $50$ from
greater to smaller deflection angles. Dashed curves: first order
approximation (\ref{oned}); solid curves: second order
approximation (\ref{higher}); dotted line: cut off for
approximation curves in the adjusted truncation; symbols:
numerical simulation results.} \label{chopping}
\end{figure}

We end this section by repeating our earlier assertion that the
perturbation theory becomes more accurate with increasing
temperature. In fact, the series expansion developed in the
Appendix is a perturbation series about the limit of infinite
forcing, and non-zero temperature acts to regularize this series
at finite forcing. The accuracy of truncations of the series thus
improves with increasing force and with increasing temperature.
 Although the sorting capability of the system
decreases with increasing temperature (and with increasing force),
it is useful to display some higher temperature results
explicitly. We do so in the one-dimensional case. By continuing
the iteration method introduced in the Appendix, we write
successive approximations to the average velocity as
\begin{equation}
v^{(U)}=U+2\pi \sum_{n=1}^N v_{n}, \label{vseries}
\end{equation}
with
\begin{eqnarray}
v_1 &=& -\frac{\tilde U}{2({\tilde U}^2+\T^2)} \nonumber\\
v_2 &=& -\frac{{\tilde U}^3-5 {\tilde U} \T^2}{8({\tilde U}^2+\T^2)^2
({\tilde U}^2+4\T^2)}\nonumber\\
v_3 &=& -\frac{{\tilde U}({\tilde U}^4-24 {\tilde U}^2\T^2+23
\T^4)}{16({\tilde U}^2+\T^2)^3({\tilde U}^4+13 {\tilde U}^2
\T^2+36 \T^4)}. \label{vapprox}
\end{eqnarray}
Here we write ${\tilde U}=U/(2\pi)$ to keep the formulas simple;
note that $v_1$ and $v_2$ have already appeared in
Eq.~(\ref{higher}). Figure~\ref{temp} show $v/U$ vs $\T$ at
forcing value $U=3$ as the number of terms retained in the series
is increased.
\begin{figure}
\centering \epsfig{figure=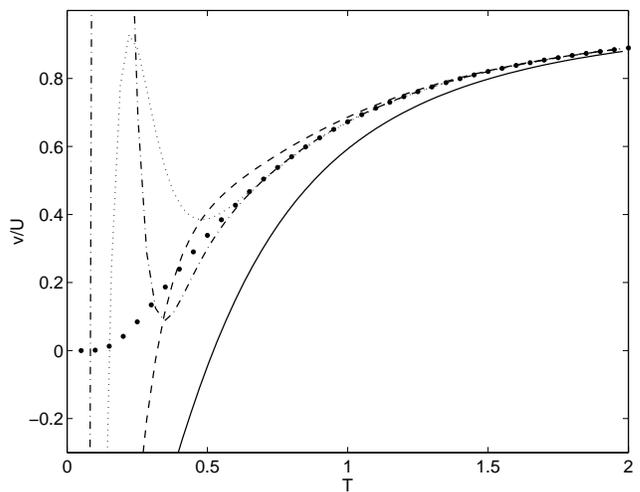,width=8.5cm}
\caption{ Results for $v/U$ for the one-dimensional potential with
forcing $U=3$, plotted as a function of temperature $\T$. Symbols
show exact values; the approximation curves use $v^{(N)}$ for $v$,
as defined in equation (\ref{vseries}), with coding: $N=1$
(solid), $N=2$ (dashed), $N=3$ (dotted), $N=4$ (dash-dotted). }
\label{temp}
\end{figure}
Note that at high temperatures, even the simple first order
($N=1$) truncation already gives accurate results. At lower
temperatures, retaining higher order terms in the approximations
gives results which are closer to the exact values. As the
temperature approaches zero for this $U$ value, the approximations
all fail; this effect depends on the forcing $U$, and at higher
values of $U$ all the approximations remain accurate even at
$\T=0$, as noted following Eq.~(\ref{higher}).

\section{Results for random potentials}
\label{resultsr} We next test our approach for random potentials,
for which the first order approximation to the velocity is given
in general in Eq.~(\ref{formula}) and Eq.~(\ref{randQ}), and in
one dimension in Eq.~(\ref{1DrandFDC1}). The second order
iteration for a one-dimensional random potential is shown in
Eq.~(\ref{secondorder}).

Recall that sorting is not possible in an isotropic random potential.
We must therefore choose an anisotropic potential.  In particular, we
consider a two-dimensional
(separable) potential generated by adding together independent,
zero-mean modulations in the $x$ and $y$ directions,
\begin{equation}
V(x,y)=V_1(x)+V_2(y), \label{27}
\end{equation}
with correlation functions given by
\begin{eqnarray}
\overline{V_1(x^\prime)V_1(x^\prime+x)}&=&E_1(x), \nonumber\\
\overline{V_2(y^\prime)V_2(y^\prime+y)}&=&E_2(y),  \nonumber\\
\overline{V_1(x^\prime)V_2(y^\prime)}&=&0.
\end{eqnarray}
The energy spectrum of this potential
has the form
\begin{equation}
\hat{E}(\kv)=2\pi\, \d(k_y)\hat{E}_1(k_x)+2\pi\, \d(k_x)\hat{E}_2(k_y),
\label{specex}
\end{equation}
where $\hat{E}_1$ and $\hat{E}_2$ are the one-dimensional spectra
given by the Fourier transforms of $V_1$ and $V_2$.
Using spectrum (\ref{specex}) in the formula (\ref{formula})
yields the average velocity components
\begin{eqnarray}
v_x &=& F\cos\th - \frac{1}{2\pi}\int_{-\infty}^\infty d k\, \frac{k^4
\hat{E}_1(k) F \cos \th}{\T^2 k^4+ k^2 F^2 \cos^2 \th}, \nonumber\\
v_y &=& F\sin\th - \frac{1}{2\pi}\int_{-\infty}^\infty d k\, \frac{k^4
\hat{E}_2(k) F \sin \th}{\T^2 k^4+ k^2 F^2 \sin^2 \th}.\label{32}
\end{eqnarray}
We note that these formulas imply that the average velocity component
perpendicular to $\Fv$ is non-zero in general (even if
$\hat{E}_1=\hat{E}_2$):
\begin{eqnarray}
v_\perp &=& -v_x \sin\th+v_y \cos \th \nonumber\\
&=& {F \cos\th \sin\th}\frac{1}{2\pi}\int_{-\infty}^\infty dk\,
k^4 \left[\frac{\hat{E}_1(k)}{\T^2 k^4+ k^2 F^2 \cos^2
\th} \right. \nonumber\\
&&\left. \hspace{0.2cm} -\frac{\hat{E}_2(k)}{\T^2 k^4+ k^2 F^2 \sin^2 \th}\right].
\label{33}
\end{eqnarray}
The separable potential (\ref{27}) is special because it reduces
the two-dimensional sorting problem to motion in uncoupled
one-dimensional potentials in the $x$ and $y$ directions.
The results~(\ref{32}) for the two-dimensional case
follow from Eq.~(\ref{1DrandFDC1})
by noting that $v_x=v(F\cos\th)$ and $v_y=v(F\sin\th)$.

A comparison between the first-order adjusted truncation results and
numerical simulations for a two-dimensional separable potential with
correlation functions
\begin{equation}
E_1(x) = \frac{\varepsilon}{\sqrt{2\pi}\lambda} e ^{-x^2/2\lambda^2},
\quad E_2(y) = \frac{\varepsilon}{\sqrt{2\pi}\lambda} e ^{-y^2/2\lambda^2},
\end{equation}
is presented in Fig.~\ref{fig6}. While the agreement is not as
good as in the periodic case, the theory clearly captures the
numerical behavior rather well, and the effect of the order of the
approximation is again manifest.

\begin{figure}
\centering
 \epsfig{figure=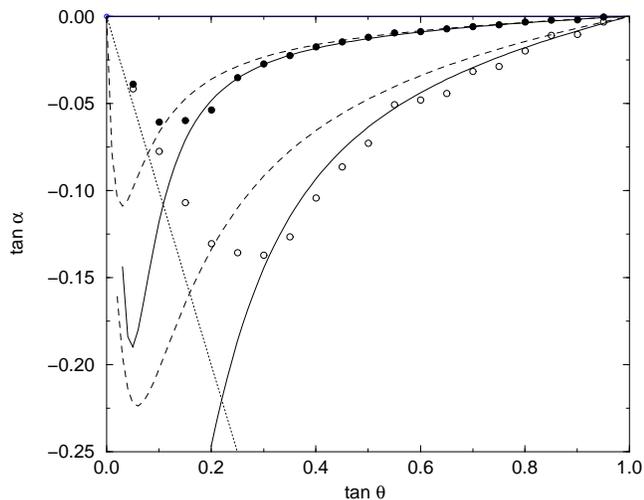,width=8.5cm}
\caption{ Deflection angles in the two-dimensional separable
random potential with Gaussian energy spectrum in each direction
($\varepsilon=5$, $\lambda=4$), temperature $\T=0.2$ and external
forcing $F=1$ (lower results) and $F=2$ (upper results). The
dashed curves show the first-order approximation, and the solid
curves the second-order approximation. The adjusted truncation
cuts off both approximations at the dotted line. The numerical
simulation results are shown as symbols.} \label{fig6}
\end{figure}

\section{Conclusion}
\label{conclusion} In summary, we have derived approximate
analytic or quadrature forms for the average velocity of particles
moving in a periodic potential or random potential. The average
velocity in general deviates from that of the applied constant
external force, the deviation depending on some particle
characteristic such as size.  This can then be used to sort
particles that differ in this characteristic. A systematic
perturbation series valid for large external forces and/or high
temperatures is shown to capture the behavior observed in
experiments and numerical simulations in physically interesting
parameter regimes when the angle between the external force and
the crystallographic $x$ axis (or $y$ axis) is relatively large,
e.g., when the force lies near the crystal diagonal, even when the
perturbation series is truncated at low orders.  The results are
not nearly as good when the force lies near one of the crystal
axes, because the truncated perturbation series can then lead to
unphysical negative velocity components. We have proposed an
adjustment to the simple perturbation expansion whereby negative
velocity components are set to zero, and have shown that this
adjusted truncation scheme leads to very good agreement with
numerical simulation results even for low temperatures.

Further directions in this work are plentiful, and here we list
just a few.   A first direction would be a generalization to
reduced-symmetry potentials.  For example, in~\cite{SPIE} we
considered a potential of the form
\begin{equation}
V(x,y)=\cos(2\pi x)\cos\left(\frac{2\pi y}{\sqrt{2}}\right),
\end{equation}
which has different length scales in the $x$ and $y$ directions.
Another extension would be to the approximate calculation of the
diffusion tensor
 \begin{equation}
D_{i j}= \lim_{t\rightarrow
\infty}\frac{d}{dt}\left<\left(x_i-\left<x_i\right>\right)
\left(x_j-\left<x_j\right>\right)\right>,
\end{equation}
(see, e.g., Eq.~(4.24) of \cite{BouchaudGeorges}), in particular
to compare values for diffusion in the transverse and parallel
directions to the direction of transport \cite{SPIE}. Reimann et
al.~\cite{Reimann} supply an exact expression for the effective
diffusion in one-dimensional problems, which should also be
applicable to the two-dimensional case with a separable potential.
These and other related investigations are in progress.

\section*{Acknowledgements}
This work is supported in part by Science Foundation Ireland under
Investigator Award 02/IN.1/IM062 (JPG), by the MCyT (Spain) under
project BFM2003-07850 (JMS, AML), and by the US National Science
Foundation under grant No. PHY-0354937 (KL).

\appendix
\section{Iteration}
In this Appendix we outline a systematic approximation scheme that leads
to our main results in Sec.~\ref{approximate}, starting from the
Fokker-Planck equation (\ref{FPE}).
We call $\uv(\xv)=-\nabla V(\xv)$ for convenience.
The case where $\uv$ is a Gaussian random field perturbing the
strong external bias $\Fv$ has been examined in, for example,
Sec. 4.2.2 of~\cite{BouchaudGeorges}. In our case the field
$\uv$ may be random or periodic, and we mimic the derivation of
the systematic expansion of~\cite{BouchaudGeorges} to determine our
results for the long-time average velocity.

Defining the (spatial) Fourier transform $\hat c$ of the
concentration field $c$ by
\begin{equation}
\hat c(\kv,t)=\int d\xv\, e^{-i \kv\cdot\xv}c(\xv,t), \label{A2}
\end{equation}
the Fokker-Planck equation is transformed to the equation
\begin{equation}
\frac{\p \hat c}{\p t}+i \kv\cdot\Fv \hat c+\frac{i}{(2\pi)^2}\int
d\pv \, \kv\cdot{\hat \uv}(\pv)\hat c(\kv-\pv,t)+k^2 \T \hat
c(\kv,t)=0, \label{A3}
\end{equation}
with initial condition $\hat c(\kv,0)=1$.
Note that the Fourier transform of $\uv$ is
\begin{equation}
\hat \uv(\pv)=-i \pv \hat V(\pv) \label{A3a}.
\end{equation}
As a consequence we have
\begin{equation}
\hat \uv(\mathbf{0})=\zv, \label{A3b}
\end{equation}
which we will use below.

We also perform a Laplace transform in time, defining
\begin{equation}
\bar c(\kv,s)=\int_0^{\infty}dt\, e^{-s t}\hat c(\kv,t).\label{A4}
\end{equation}
After Laplace transforming Eq.~(\ref{A3}) and using the
initial condition, our goal becomes the solution of the integral
equation
\begin{equation}
\bar c(\kv,s) = P_s(\kv)-\frac{i}{(2\pi)^2}P_s(\kv)\int d\pv\,
\kv\cdot\hat \uv(\pv)\bar c(\kv-\pv,s), \label{A5}
\end{equation}
where the propagator (or ``free Green function''
\cite{BouchaudGeorges}) is defined as
\begin{equation}
P_s(\kv)=\frac{1}{s+k^2\T+i \kv\cdot\Fv}. \label{A6}
\end{equation}
The following limiting behavior of the propagator will be
important later:
\begin{equation}
\lim_{s\rightarrow 0} s P_s(\kv) = \left\{
\begin{array}{c c}
0 & \text{ if }\kv \neq \zv \\
1 & \text{ if }\kv  =\zv.
\end{array}
\right.\label{A7}
\end{equation}

If the exact solution $\bar c(\kv,s)$ of Eq.~(\ref{A5}) could be
found, then the thermal-averaged velocity defined in
Eq.~(\ref{vel}) can be determined using standard limiting theorems
for Laplace transforms as
\begin{equation}
\left<\vv\right>=\lim_{s\rightarrow 0} i s^2 \left.\frac{\p \bar
c}{\p \kv}\right|_{\kv=\zv} \label{A8}.
\end{equation}
However, as Eq.~(\ref{A5}) is not in general exactly
solvable, we seek instead an approximate solution for $\bar
c(\kv,s)$. A standard approach to approximating the solution of an
integral equation is by iterating; i.e., first setting $\bar
c(\kv,s)=P_s(\kv)$ (neglecting the second term on the right hand
of Eq.~(\ref{A5})), then substituting this approximate solution
into (\ref{A5}) to obtain an updated solution, etc. After two
iterations this procedure yields:
\begin{eqnarray}
\bar c(\kv,s) &=& P_s(\kv) -\frac{i}{(2\pi)^2} P_s(\kv) \int
d\pv\,
\kv\cdot\hat\uv(\pv) P_s(\kv-\pv) \nonumber\\
&&-\frac{1}{(2\pi)^4}P_s(\kv)\int d\pv\,
\kv\cdot\hat\uv(\pv)P_s(\kv-\pv)\nonumber\\
&&\times \int d\qv\,
(\kv-\pv)\cdot\hat\uv(\qv)P_s(\kv-\pv-\qv)+\ldots, \nonumber\\
&&
\label{A9}
\end{eqnarray}
with the dots signifying the further terms in the formal iteration
series.
Successive powers of the propagator appear at each iteration,
and therefore the
approximation is assumed to be better for large $F$ and large $\T$.

Consider now the velocity $\left<\vv\right>$ in Eq.~(\ref{A8})
which results from using the approximation that terminates the
series (\ref{A9}) by neglecting the dots.  First, we differentiate
$\bar c$ with respect to the component $k_j$ and evaluate at
$\kv=\zv$:
\begin{eqnarray}
\left. \frac{\p\bar c}{\p k_j}\right|_{\kv=\zv} &=& \frac{-i}{s^2}
F_j -\frac{i}{(2\pi)^2} P_s(\zv) \int d\pv\,
\hat{u}_j(\pv) P_s(-\pv) \nonumber\\
&&-\frac{1}{(2\pi)^4}P_s(\zv)\int d\pv\,
\hat{u}_j(\pv)P_s(-\pv) \nonumber\\
&&\times \int d\qv\,
(-\pv)\cdot\hat\uv(\qv)P_s(-\pv-\qv) \label{A10}.
\end{eqnarray}
Following (\ref{A8}), we now multiply this equation by $i s^2$ and
examine the limit as $s\rightarrow 0$, and find that the first
term of (\ref{A10}) yields $F_j$, i.e. the unmodified influence of
the external force on the velocity. The second term of (\ref{A10})
reduces in the limit to
\begin{equation*}
\frac{1}{(2\pi)^2}  \int d\pv\, \hat{u}_j(\pv)
\left(\lim_{s\rightarrow 0}s P_s(-\pv)\right),
\end{equation*}
which vanishes, since $\lim_{s\rightarrow 0} s P_s(-\pv)=0$ unless
$\pv=\zv$ [by (\ref{A7})], and $\hat \uv(\zv)=\zv$ [by
(\ref{A3b})]. To evaluate the limit of the third term of
(\ref{A10}), we write it as
\begin{equation}
\lim_{s\rightarrow 0}\left\{\frac{i}{(2\pi)^4} s \int d\pv\,
\hat{u}_j(\pv)P_s(-\pv)\int d\qv\,
\pv\cdot\hat\uv(\qv)P_s(-\pv-\qv)\right\}, \label{A11}
\end{equation}
and note that
\begin{eqnarray}
&&\lim_{s\rightarrow 0}s P_s(-\pv)P_s(-\pv-\qv) \nonumber\\
&&= \displaystyle{\lim_{s\rightarrow 0}}
\frac{\displaystyle s}{\displaystyle
\left[s+p^2\T-i\pv\cdot\Fv\right]
\left[s+(\pv+\qv)^2\T-i(\pv+\qv)\cdot\Fv\right] }   \nonumber \\
&& = 0 \text{, unless }\pv=\zv\text{ or }\pv+\qv=\zv.
\end{eqnarray}
Since $\hat{u}_j(\pv)$ is zero if $\pv=\zv$ [by Eq.~(\ref{A3b})],
we conclude that the only non-zero contribution of
(\ref{A11}) occurs when $\qv=-\pv$, giving the value
\begin{equation}
\frac{i}{(2\pi)^4}\int d\pv \, \hat{u}_j(\pv)\frac{1}{\T p^2-i
\pv\cdot \Fv}\, \pv\cdot \hat{\uv}(-\pv) \label{A12}.
\end{equation}
Together with the contribution from the first term of (\ref{A10}),
we thus have the approximation
\begin{equation}
\left<\vv\right> \approx \Fv + \frac{i}{(2\pi)^4}\int d\pv\, \hat
\uv(\pv) \frac{1}{\T p^2-i \pv\cdot \Fv}\,\pv\cdot \hat{\uv}(-\pv)
\label{A13}.
\end{equation}
Equation (\ref{formula}) follows on replacing $\hat \uv$ from
equation (\ref{A3a}); the random potential case then requires a
further ensemble average over the disorder to yield equation
(\ref{randQ}).

\end{document}